\documentclass[a4paper,fleqn,usenatbib,letter]{mnras}

\usepackage[T1]{fontenc}
\usepackage{ae,aecompl}

\usepackage{graphicx}	% Including figure files
\usepackage{amsmath}	% Advanced maths commands
\usepackage{amssymb}	% Extra maths symbols

\title[Post-outburst observations of V346\,Normae]{V346\,Normae: First post-outburst observations of an FU\,Orionis star
\thanks{Based on observations made with ESO telescopes 
at the Paranal Observatory under programme IDs 
075.C-0625(A,B,C),
079.C-0613(A,B,C,D), 
179.B-2002(A,B,C), and
095.C-0765(A,B).}
}

\author[S.~Kraus et al.]{S.~Kraus,$^{1}$  %\thanks{E-mail: skraus@astro.ex.ac.uk}, 
A.~Caratti o Garatti,$^{2}$
R.~Garcia-Lopez,$^{2}$
A.~Kreplin,$^{1}$
A.~Aarnio,$^{3}$
\newauthor J.D.~Monnier,$^{3}$
T.~Naylor,$^{1}$
G.~Weigelt$^{4}$
\\
$^{1}$~School of Physics, Astrophysics Group, University of Exeter, Stocker Road, Exeter EX4 4QL, UK\\
$^{2}$~Dublin Institute for Advanced Studies, School of Cosmic Physics, Astronomy \& Astrophysics Section, 31 Fitzwilliam Place, Dublin 2, Ireland\\
$^{3}$~Department of Astronomy, University of Michigan, 311 West Hall, 1085 South University Ave, Ann Arbor, MI 48109, USA\\
$^{4}$~Max-Planck-Institut f\"ur Radioastronomie, Auf dem H\"ugel 69, 53121 Bonn, Germany
}

\date{Accepted 2016-06-21. Received 2016-05-27; in original form 2016-04-18}

\pubyear{2016}

\begin{document}
\label{firstpage}
\pagerange{\pageref{firstpage}--\pageref{lastpage}}
\maketitle

\begin{abstract}
  During their formation phase stars gain most of their mass in violent episodic 
  accretion events, such as observed in FU\,Orionis (FUor) and EXor stars.
  V346\,Normae is a well-studied FUor that underwent a strong outburst 
  beginning in $\sim$\,1980.  Here, we report photometric and spectroscopic
  observations which show that the visual/near-infrared brightness has decreased dramatically between 
  the 1990s and 2010 (${\Delta}R\approx10.9^{\rm m}$, ${\Delta}J\approx7.8^{\rm m}$, ${\Delta}K\approx5.8^{\rm m}$).
The spectral properties of this fading event cannot be explained with variable extinction alone, 
    but indicate a drop in accretion rate by 2-3 orders of magnitude, marking the first time 
    that a member of the FUor class has been observed to switch to a very low accretion phase.
    Remarkably, in the last few years (2011-2015) V346\,Nor has brightened again
    at all near-infrared wavelengths, indicating the onset of a new outburst event.
    The observed behaviour might be consistent with the {\it clustered luminosity bursts}
    that have been predicted by recent gravitational instability and fragmentation models
    for the early stages of protostellar evolution.
  Given V346\,Nor's unique characteristics (concerning outburst duration,
  repetition frequency, and spectroscopic diagnostics),
  our results also highlight the need for revisiting the FUor/EXor classification scheme.
\end{abstract}


%%%%%%%%%%%%%%%%%%%%%%%%%%%%%%%%%%%%%%%%%%%%%%%%%%

%%%%%%%%%%%%%%%%% BODY OF PAPER %%%%%%%%%%%%%%%%%%

\section{Introduction}

FU\,Orionis stars (FUors) are pre-main-sequence stars that undergo 
optical outbursts, interpreted as extremely active phases of mass accretion
($\sim10^{-4}\,$M$_{\sun}$\,yr$^{-1}$).
These sources are classified by the observation of outbursts that 
increase the optical/infrared brightness by up to 6\,mag for several decades
as well as their spectroscopic characteristics, which include
CO overtone bandhead $2.3$\,$\mu$m absorption
and P\,Cygni H$\alpha$~profiles that point towards strong outflow activity
\citep[for a recent review see][]{aud14}.
Only about a dozen stars are generally considered to belong to the FUor
class, with a few more dozen FUor candidates.
Besides their extremely high accretion luminosities, FUor outbursts are 
characterized by their very long outburst durations \citep[estimated to $\sim10^2$...$10^3$\,years;][]{har96} 
- in fact the commonly accepted FUor members are still in outburst since their
discovery decades ago, including the prototype FU\,Orionis itself.
The long outburst durations are also used to separate FUors
from the EXors, which are eruptive stars of shorter 
outburst decay time (months to few years) and with 
lower accretion rates \citep[$\sim10^{-6}\,$M$_{\sun}$\,yr$^{-1}$,][]{her89}.

V346\,Normae is in the Sa~187 cloud at a distance of $700$\,pc 
and is generally considered to be a genuine member of the FUor class
\citep{pru93,rei97a}.
Outburst activity of V346~Nor was first discovered in 1983 when \citet{gra85}
reported the appearance of a star-like object in the visible
($V=16$\,mag).
\citet{abr04} compiled the light curve and concluded 
that the near-infrared brightness was increasing
continuously throughout the 1990s.
The FUor classification is also based on the detection of strong Li\,I absorption, 
a P Cygni-like H$\alpha$ profile, and the detection of strong 
water vapour absorption around 1.9\,$\mu$m \citep{rei85}.

Here, we report photometric observations (Sect.~\ref{sec:observations}) which show that V346\,Nor 
has experienced a dramatic drop in visual/infrared excess, indicating
that the object has switched from a high-accretion to a low-accretion phase.
We discuss our observed dramatic variability in Sect.~\ref{sec:results} and 
close with a summary and discussion on the broader implications of our study in Sect.~\ref{sec:conclusions}.

\section{Observations}
\label{sec:observations}

\begin{figure}
  \centering
  \hspace{4mm}
  \includegraphics[angle=270,scale=0.28]{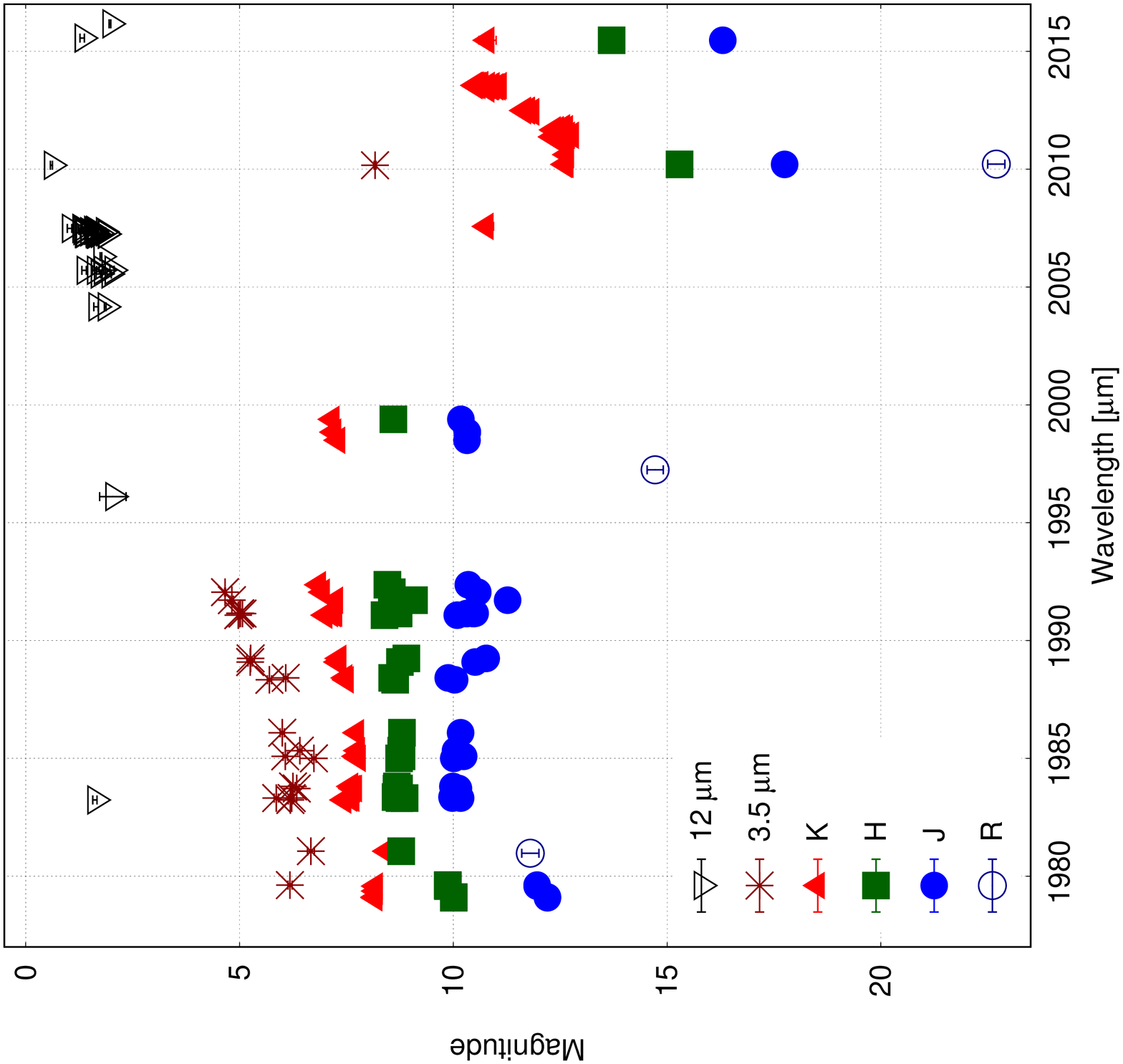}
  \includegraphics[angle=0,scale=0.42]{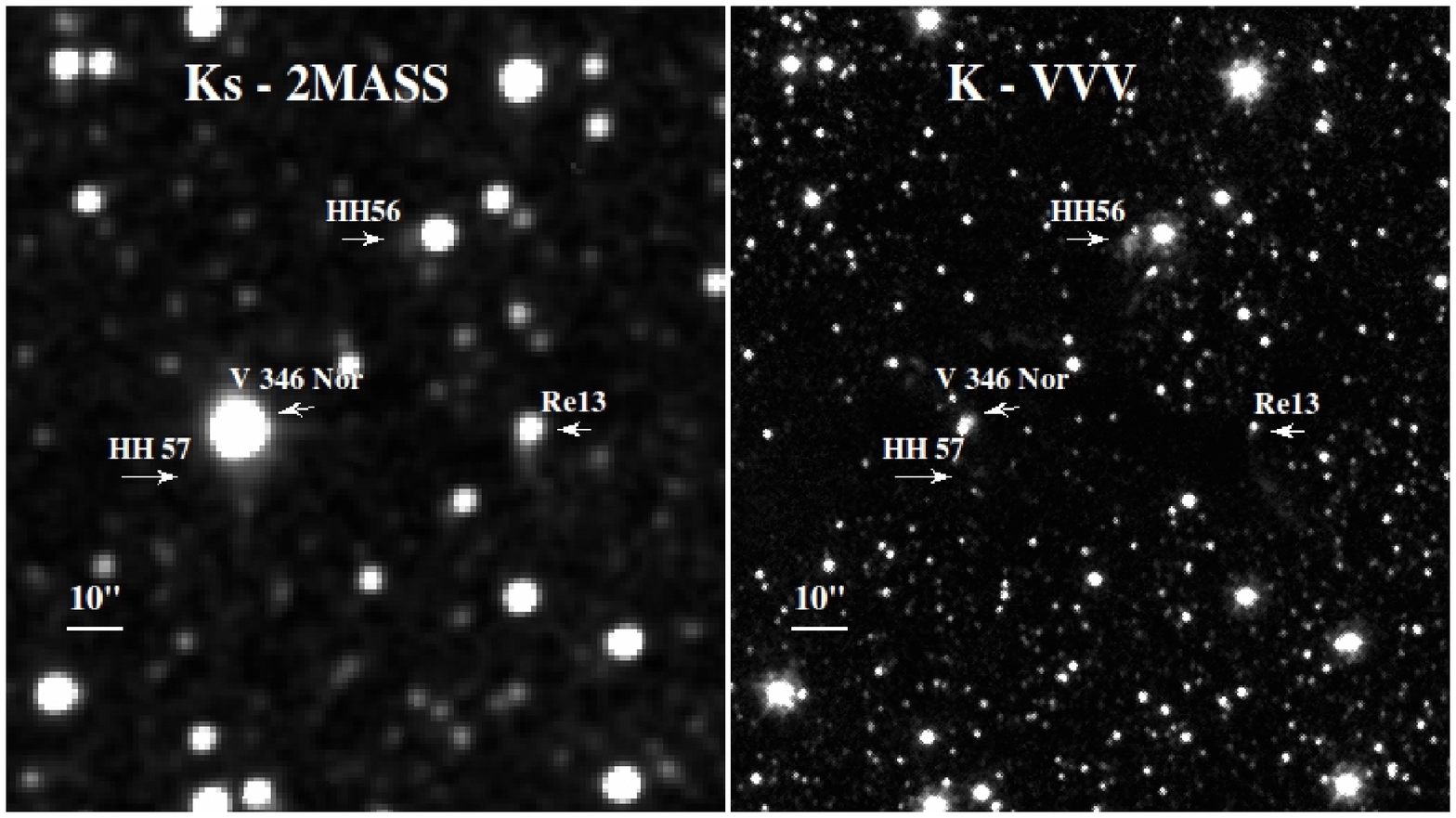}
  \vspace{-2mm}
  \caption{
    Top: Lightcurve of V346\,Nor, including R/J/H/K-band and 3.5 and 12\,$\mu$m.
    Bottom: The field around V346\,Nor in outburst (left column; 1999, 2MASS) and in the low-accretion phase (right column; 2011, VVV).
    \label{fig:photometry}}
\end{figure}

\begin{figure}
  \centering
  \includegraphics[angle=270,scale=0.35]{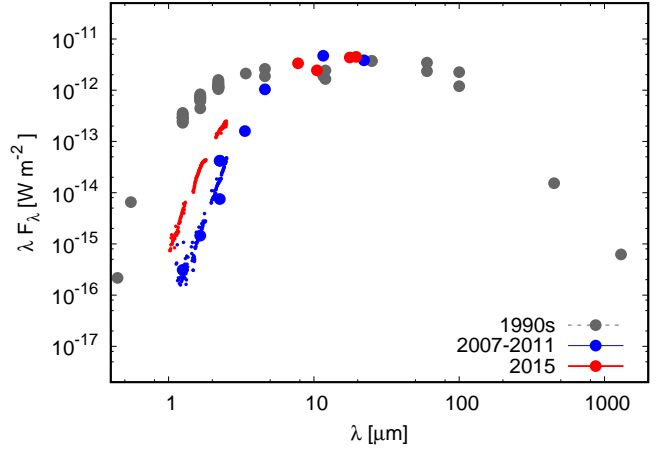} 
  \caption{
    SED measured at three epochs (1990s, 2007-2011, 2015).
%      overplotted with the interpolation curve that we use to estimate the change in 
 %     near- to mid-infrared luminosity between the different epochs.
  }
  \label{fig:SED}
\end{figure}

\begin{table}
  \centering
 \caption{Photometry on V346\,Nor}
 \label{tab:photometry}
 \begin{tabular}{ccccc}
  \hline
   Epoch & Band & Magnitude/Flux & Ref.\\
  \hline
1980-12-22  & R & $11.8$                    & mon98\\
1997-03-30   & R & $14.72\pm0.19$   &  las08\\
\hline
1979-02-01  & J &  $12.20 \pm 0.08$ & mol93\\
1979-02-11  & J &  $12.20 \pm 0.05$ & gra85\\
1979-08-??  & J &  $11.96 \pm 0.08$ & rei83\\
1983-04-25  & J &  $10.17 \pm 0.03$ & rei85\\
1983-04-25  & J &  $9.98 \pm 0.05$ & gra85\\
1983-05-10  & J &  $9.97 \pm 0.05$ & gra85\\
1983-09-19  & J & $10.08 \pm 0.05$ & gra85\\
1983-09-21  & J &  $10.12 \pm 0.05$ & gra85\\
1983-10-23  & J &  $9.99 \pm 0.05$ & gra85\\
1985-01-01  & J &  $10.01 \pm 0.02$ & mol93\\
1985-02-01  & J &  $10.24 \pm 0.02$ & mol93\\
1985-05-01  & J &  $10.05 \pm 0.02$ & mol93\\
1986-02-01  & J &  $10.17 \pm 0.02$ & mol93\\
1988-05-01  & J &  $10.03 \pm 0.02$ & mol93\\
1988-06-01  & J &  $9.88 \pm 0.02$ & mol93\\
1989-02-01  & J &  $10.51 \pm 0.02$ & mol93\\
1989-03-30  & J &  $10.77 \pm 0.03$ & mol93\\
1991-01-27  & J &  $10.09 \pm 0.01$ & mol93\\
1991-02-22  & J &  $10.31 \pm 0.02$ & mol93\\
1991-02-22  & J &  $10.47 \pm 0.01$ & mol93\\
1991-02-26  & J &  $10.51 \pm 0.01$ & mol93\\
1991-09-18  & J &  $11.27 \pm 0.04$ & mol93\\
1992-01-20  & J &  $10.57 \pm 0.02$ & mol93\\
1992-05-13  & J &  $10.35 \pm 0.01$ & pru93\\
1998-07-01  & J &  $10.318 \pm 0.060$ & den05\\
1998-11-03  & J &  $10.327 \pm 0.070$ & den05\\
1999-05-20  & J &  $10.178 \pm 0.025$ & cut03\\
2010-03-16  & J &  $17.749 \pm 0.076$ & min10\\
2015-06-20 &  J &  $16.3 \pm 0.2$ & XSHOOTER\\
\hline
1979-02-01  & H &  $10.01 \pm 0.07$ & mol93\\
1979-02-11  & H &   $10.01 \pm 0.05$ & gra85\\
1979-08-??  & H &   $9.87 \pm 0.03$ & rei83\\
1981-01-23  & H &   $8.78 \pm 0.13$ & rei85\\
1983-04-25  & H &   $8.74 \pm 0.02$ & rei85\\
1983-04-25  & H &   $8.86 \pm 0.05$ & gra85\\
1983-05-10  & H &   $8.65 \pm 0.05$ & gra85\\
1983-09-19  & H &   $8.67 \pm 0.05$ & gra85\\
1983-09-21  & H &   $8.73 \pm 0.05$ & gra85\\
1983-10-23  & H &   $8.68 \pm 0.05$ & gra85\\
1985-01-01  & H &   $8.73 \pm 0.02$ & mol93\\
1985-02-01  & H &   $8.79 \pm 0.02$ & mol93\\
1985-05-01  & H &   $8.78 \pm 0.02$ & mol93\\
1986-02-01  & H &   $8.80 \pm 0.02$ & mol93\\
1988-05-01  & H &   $8.64 \pm 0.02$ & mol93\\
1988-06-01  & H &   $8.57 \pm 0.02$ & mol93\\
1989-02-01  & H &   $8.75 \pm 0.02$ & mol93\\
1989-03-30  & H &   $8.90 \pm 0.02$ & mol93\\
1991-01-27  & H &   $8.39 \pm 0.01$ & mol93\\
1991-02-22  & H &   $8.59 \pm 0.02$ & mol93\\
1991-02-22  & H &   $8.70 \pm 0.01$ & mol93\\
1991-02-26  & H &   $8.72 \pm 0.02$ & mol93\\
1991-09-18  & H &   $9.08 \pm 0.03$ & mol93\\
1992-01-20  & H &   $8.56 \pm 0.01$ & mol93\\
1992-05-13  & H &   $8.46 \pm 0.01$ & pru93\\
1999-05-20  & H &   $8.599 \pm 0.031$ & cut03\\
2010-03-16  & H &   $15.289 \pm 0.017$ & min10\\
2015-06-20 &  H &  $13.7 \pm 0.2$ & XSHOOTER\\
\hline
1983-04-25  & K &   $7.64 \pm 0.01$ & rei85\\
1983-04-25  & K &   $7.59 \pm 0.05$ & gra85\\
1983-05-10  & K &   $7.56 \pm 0.05$ & gra85\\
\hline
 \end{tabular}
\end{table}

\begin{table}
  \centering
 \contcaption{}
 \label{tab:photometry2}
 \begin{tabular}{ccccc}
  \hline
   Epoch & Band & Magnitude/Flux & Ref.\\
  \hline
1983-09-19  & K &   $7.66 \pm 0.05$ & gra85\\
1983-09-21  & K &   $7.70 \pm 0.05$ & gra85\\
1983-10-23  & K &   $7.61 \pm 0.10$ & gra85\\
1985-01-01  & K &   $7.79 \pm 0.02$ & mol93\\
1985-02-01  & K &   $7.73 \pm 0.02$ & mol93\\
1985-05-01  & K &   $7.77 \pm 0.02$ & mol93\\
1986-02-01  & K &   $7.75 \pm 0.02$ & mol93\\
1988-05-01  & K &   $7.51 \pm 0.02$ & mol93\\
1988-06-01  & K &   $7.48 \pm 0.02$ & mol93\\
1989-02-01  & K &   $7.31 \pm 0.02$ & mol93\\
1989-03-30  & K &   $7.34 \pm 0.01$ & mol93\\
1991-01-27  & K &   $7.01 \pm 0.01$ & mol93\\
1991-02-22  & K &   $7.14 \pm 0.02$ & mol93\\
1991-02-22  & K &   $7.21 \pm 0.01$ & mol93\\
1991-02-26  & K &   $7.24 \pm 0.01$ & mol93\\
1991-09-18  & K &   $7.26 \pm 0.03$ & mol93\\
1992-01-20  & K &   $6.95 \pm 0.01$ & mol93\\
1992-05-13  & K &   $6.86 \pm 0.01$ & pru93\\
1998-07-01  & K &   $7.312 \pm 0.070$ & den05\\
1998-11-03  & K &   $7.215 \pm 0.070$ & den05\\
1999-05-20  & K &   $7.176 \pm 0.021$ & cut03\\
2007-07-28  & K &   $10.78 \pm 0.16$ & SINFONI\\
2010-03-15  & K &   $12.615 \pm 0.003$ & min10\\
2010-03-16  & K &   $12.649 \pm 0.010$ & min10\\
2010-08-14  & K &   $12.662 \pm 0.004$ & min10\\
2011-05-13  & K &   $12.33 \pm 0.05$ & min10\\
2011-05-14  & K &   $12.38 \pm 0.06$ & min10\\
2011-05-16  & K &   $12.33 \pm 0.06$ & min10\\
2011-05-17  & K &   $12.39 \pm 0.06$ & min10\\
2011-05-30  & K &   $12.59 \pm 0.05$ & min10\\
2011-06-09  & K &   $12.68 \pm 0.06$ & min10\\
2011-08-18  & K &   $12.38 \pm 0.05$ & min10\\
2011-08-23  & K &   $12.43 \pm 0.08$ & min10\\
2011-08-31  & K &   $12.492 \pm 0.05$ & min10\\
2011-09-01  & K &   $12.35 \pm 0.06$ & min10\\
2011-09-01  & K &   $12.513 \pm 0.05$ & min10\\
2011-09-02  & K &   $12.42 \pm 0.06$ & min10\\
2011-09-19  & K &   $12.635 \pm 0.05$ & min10\\
2011-09-20  & K &   $12.55 \pm 0.06$ & min10\\
2011-10-02  & K &   $12.57 \pm 0.06$ & min10\\
2012-06-05  & K &   $11.85 \pm 0.05$ & min10\\
2012-06-05  & K &   $11.84 \pm 0.05$ & min10\\
2012-06-20  & K &   $11.75 \pm 0.05$ & min10\\
2012-06-21  & K &   $11.75 \pm 0.05$ & min10\\
2012-06-27  & K &   $11.65 \pm 0.06$ & min10\\
2012-06-27  & K &   $11.66 \pm 0.06$ & min10\\
2012-06-30  & K &   $11.74 \pm 0.06$ & min10\\
2013-05-27  & K &   $10.8 \pm 0.0.1$ & min10\\
2013-06-27  & K &   $10.77 \pm 0.05$ & min10\\
2013-06-28  & K &   $11.06 \pm 0.05$ & min10\\
2013-06-30  & K &   $11.01 \pm 0.05$ & min10\\
2013-07-01  & K &   $10.90 \pm 0.05$ & min10\\
2013-07-02  & K &   $10.88 \pm 0.05$ & min10\\
2013-07-14  & K &   $10.92 \pm 0.06$ & min10\\
2013-07-17  & K &   $11.09 \pm 0.06$ & min10\\
2013-07-20  & K &   $10.52 \pm 0.06$ & min10\\
2013-07-21  & K &   $10.51 \pm 0.06$ & min10\\
2013-07-21  & K &   $10.59 \pm 0.06$ & min10\\
2013-07-22  & K &   $10.81 \pm 0.06$ & min10\\
2013-07-25  & K &   $10.55 \pm 0.09$ & min10\\
2013-07-28  & K &   $10.66 \pm 0.08$ & min10\\
2013-07-31  & K &   $10.65 \pm 0.05$ & min10\\
2015-06-20 &  K &   $10.8 \pm 0.2$ & XSHOOTER\\
\hline
 \end{tabular}
\end{table}

\begin{table}
  \centering
 \contcaption{}
 \label{tab:photometry3}
 \begin{tabular}{ccccc}
  \hline
   Epoch & Band & Magnitude/Flux & Ref.\\
  \hline
1979-08-??  & $3.5\,\mu$m &  $6.18 \pm  0.03$ & rei83\\
1981-01-23  & $3.6\,\mu$m &  $6.57 \pm 0.08$ & rei85\\
1983-03-26  & $3.5\,\mu$m &  $6.19 \pm 0.05$ & gra85\\
1983-04-25  & $3.5\,\mu$m &  $5.86 \pm 0.05$ & rei85\\
1983-04-25  & $3.5\,\mu$m &  $6.22 \pm 0.05$ & gra85\\
1983-09-21  & $3.5\,\mu$m &  $6.34 \pm 0.05$ & gra85\\
1983-10-23  & $3.5\,\mu$m &  $6.25 \pm 0.05$ & gra85\\
1985-01-01  & $3.5\,\mu$m &  $6.74 \pm 0.05$ & mol93\\
1985-02-01  & $3.5\,\mu$m &  $6.07 \pm 0.05$ & mol93\\
1985-05-01  & $3.5\,\mu$m &  $6.41 \pm 0.05$ & mol93\\
1986-02-01  & $3.5\,\mu$m &  $6.00 \pm 0.05$ & mol93\\
1988-05-01  & $3.5\,\mu$m &  $5.70 \pm 0.05$ & mol93\\
1988-06-01  & $3.5\,\mu$m &  $6.08 \pm 0.05$ & mol93\\
1989-02-01  & $3.5\,\mu$m &  $5.25 \pm 0.05$ & mol93\\
1989-03-30  & $3.8\,\mu$m &  $5.26 \pm 0.09$ & mol93\\
1991-01-27  & $3.8\,\mu$m &  $4.97 \pm 0.02$ & mol93\\
1991-02-22  & $3.8\,\mu$m &  $5.01 \pm 0.02$ & mol93\\
1991-02-22  & $3.8\,\mu$m &  $5.07 \pm 0.01$ & mol93\\
1991-02-26  & $3.8\,\mu$m &  $5.02 \pm 0.02$ & mol93\\
1991-09-18  & $3.8\,\mu$m &  $4.82 \pm 0.04$ & mol93\\
1992-01-20  & $3.8\,\mu$m &  $4.66 \pm 0.01$ & mol93\\
2010-03-01 & $3.35\,\mu$m & $0.177 \pm 0.005$\,Jy & cut14\\
2010-03-01 & $4.6\,\mu$m & $1.59 \pm 0.09$\,Jy & cut14\\
%1983-04-25  & $4.6\,\mu$m &  $5.40 \pm 0.24$ & rei85\\
%1989-03-30  & $4.6\,\mu$m &  $4.70 \pm 0.20$ & mol93\\
%1991-01-27  & $4.6\,\mu$m &  $4.30 \pm 0.06$ & mol93\\
%1991-02-22  & $4.6\,\mu$m &  $4.21 \pm 0.09$ & mol93\\
%1991-02-22  & $4.6\,\mu$m &  $4.43 \pm 0.03$ & mol93\\
%1991-02-26  & $4.6\,\mu$m &  $4.30 \pm 0.06$ & mol93\\
%1991-09-18  & $4.6\,\mu$m &  $4.07 \pm 0.04$ & mol93\\
%1992-01-20  & $4.6\,\mu$m &  $4.05 \pm 0.09$ & mol93\\
%1983-03-26  & $4.8\,\mu$m &  $4.99$ & gra85\\
\hline
2010-03-01 & $11.6\,\mu$m & $18.2 \pm 0.4$\,Jy & cut14\\
2004-02-27  & $12\,\mu$m  & 7.18\,Jy & IRS\\
2005-07-21  & $12\,\mu$m  & $6.6 \pm 0.3$\,Jy & MIDI\\
2005-09-16  & $12\,\mu$m  & $9.0 \pm 0.4$\,Jy & MIDI\\
2005-09-18  & $12\,\mu$m  & $6.2 \pm 0.3$\,Jy & MIDI\\
2006-04-16  & $12\,\mu$m  & 7.87\,Jy & IRS\\
2007-03-30  & $12\,\mu$m  & $7.2 \pm 0.4$\,Jy & MIDI\\
2007-03-30  & $12\,\mu$m  & $7.1 \pm 0.4$\,Jy & MIDI\\
2007-04-09  & $12\,\mu$m  & $9.1 \pm 0.5$\,Jy & MIDI\\
2007-04-10  & $12\,\mu$m  & $9.3 \pm 0.5$\,Jy & MIDI\\
2007-04-10  & $12\,\mu$m  & $8.3 \pm 0.4$\,Jy & MIDI\\
2007-05-06  & $12\,\mu$m  & $7.4 \pm 0.4$\,Jy & MIDI\\
2007-05-28  & $12\,\mu$m  & $9.6 \pm 0.5$\,Jy & MIDI\\
2007-05-28  & $12\,\mu$m  & $9.2 \pm 0.5$\,Jy & MIDI\\
2007-06-27  & $12\,\mu$m  & $12.3 \pm 0.6$\,Jy & MIDI\\
2015-07-25  & $12\,\mu$m  & $11.7 \pm 0.5$\,Jy & VISIR\\
2010-03-01 & $22.1\,\mu$m & $28.259 \pm 0.16$\,Jy & cut14\\
\hline
 \end{tabular}
\end{table}

\begin{table}
 \centering
 \contcaption{}
 \label{tab:photometry4}
 \begin{tabular}{ccccc}
  \hline
   Epoch & Band & Magnitude/Flux & Ref.\\
  \hline
2010-03-19 & B & $<23.6$~$^{\mathrm{a}}$ & IMACS \\
2010-03-19 & V & $<23.3$~$^{\mathrm{a}}$ & IMACS \\
2010-03-19 & R & $22.7 \pm 0.2$ & IMACS \\
2015-07-25  & $7.78\,\mu$m  & $8.7 \pm 0.5$\,Jy & VISIR\\
2015-07-25  & $10.49\,\mu$m  & $8.5 \pm 0.5$\,Jy & VISIR\\
2015-07-25  & $17.65\,\mu$m  & $25.6 \pm 0.5$\,Jy & VISIR\\
2015-07-25  & $19.50\,\mu$m  & $29.1 \pm 0.5$\,Jy & VISIR\\
2016-02-29  & $12\,\mu$m  & $6.5 \pm 0.2$\,Jy & VISIR\\
\hline
 \end{tabular}
\footnotetext{Footnote}
\begin{flushleft}
  {\it Notes}~--~$^{\mathrm{a}}$~The source has not been detected in these images, therefore we provide upper limits. 
{\it References}~--~The photometric data has been taken from the following references (while new measurements are referenced with the instrument name):
      rei83: \citet{rei83}; 
      rei85: \citet{rei85}; 
      gra85: \citet{gra85}; 
      mol93: \citet{mol93}; 
      mon98: \citet{mon98};
      cut03: \citet[][2MASS]{cut03}; 
      den05: DENIS consortium (September\,2005); 
      las08: \citet[][GSC2.3.2]{las08}
      min10: \citet[][VVV]{min10};
      cut14: \citet[][WISE]{cut14}
\end{flushleft}
\end{table}

We construct the lightcurve of V346\,Nor using the following 
near- and mid-infrared imaging and spectroscopy data:
\begin{itemize}
\item VLT/SINFONI \citep[2007-07-28,][]{eis03}: 
  Integral field spectroscopy, covering the
  K-band with spectral resolution $R=4000$ and pixel sizes
  of 25 and 250\,mas/pixel.
\item Magellan/FIRE \citep[2011-03-16,][]{sim08}: Echelle spectroscopy,
  covering the J/H/K-band with $R=8000$ ({0.45\arcsec} slit).
  We also recorded J-band acquisition images on V346\,Nor
  and standard stars. These were used to check for
  slit losses that might have affected the absolute flux calibration.
\item VLT/XSHOOTER \citep[2015-06-20,][]{ver11}: Echelle mode spectroscopy,
  covering the J/H/K-band with $R=10,500$
  (slit size of {0.4\arcsec} in the near-infrared a  rm).
  Additional data was taken with a wide {5\arcsec} slit
    and used for the photometric calibration to avoid slit losses.
\item VLT/VISIR \citep[2015-07-25,][]{lag04}: Imaging in a 
  J7.9 ($7.78\pm0.55\,\mu$m), 
  SiV  ($10.49\pm0.16\,\mu$m), 
 Q1 ($17.65\pm0.83\,\mu$m), and
  Q3 ($19.50\pm0.40\,\mu$m) filter.  
\item VLT/VISIR (2016-02-29):
  Long-slit spectroscopy, covering the N-band with $R=350$.
\item VLTI/MIDI (2005-07-22/24, 2005-09-17/19,
    2007-03-31, 2007-04-10/11, 2007-05-06/29,
    2007-06-28, 2007-07-24): 
    Mid-infrared interferometry, covering the N-band with $R=30$. 
    Photometry frames were recorded both on V346\,Nor and 
    the calibrator stars, allowing us to retrieve absolute fluxes. 
\item Magellan/IMACS (2011-03-19): 
  Wide-field imaging in the B/V/R-band with 
  a pixel size of 0.11\arcsec/pixel.
\end{itemize}

We reduced the SINFONI, XSHOOTER, and VISIR data using the
standard ESO instrument pipelines (versions~2.7.0, 2.6.8, and 4.1.7, respectively),
while the MIT data reduction pipeline was used to extract the FIRE spectrum.
All spectroscopic observations were accompanied by standard star observations
(2007-07-28: HD139579; 2011-03-16: HD103125;
2015-06-20: HIP85885, HR6572;
2015-07-25, 2015-08-08: HD178345;
2016-02-29: HD149447, HD151680),
which allows us to flux-calibrate the spectra.
The MIDI data was reduced using the 
MIA+EWS software \citep[Release~2.0;][]{jaf04,lei04}
and observations on the calibrator stars HD152820, HD220704,
HD163376, HD146791, HD160668, and HD138816 were
used to extract absolute-calibrated photometry.

In this paper, we use the derived spectra to build the 
lightcurve (Figs.~\ref{fig:photometry}, top).
The line emission signatures and interferometry results
will be discussed in a future paper (Kraus et al., in prep.).

We also include photometry from
\citet{rei83,rei85,gra85,mol93}, 
\citet[][VVV]{min10}, \citet[][WISE]{cut14}, 
the USNO, GSC2, 2MASS, and DENIS catalogue, and Spitzer/IRS spectroscopy 
\citep[AORKEY\,3570688+16260864; first epoch published in][]{gre06}.
The derived photometry measurements are listed in 
Tab.~\ref{tab:photometry}, together with a compilation of
measurements from the literature.

\section{Results}
\label{sec:results}

Our lightcurve (Fig.~\ref{fig:photometry}, top) shows that the near-infrared
magnitude of V346\,Nor dropped by $\Delta K=5.8$\,mag, $\Delta J=7.8$\,mag,
and $\Delta R=10.9$\,mag between the late 1990s and 2011.
The dramatic drop in brightness is also illustrated in Fig.~\ref{fig:photometry} (bottom),
where we compare images from 1999 and 2011.
From this data, we construct the spectral energy distribution (SED) in the outburst (1990s), 
near the photometric minimum (2007-2011), and in the latest brightening phase (2015),
as shown in Fig.~\ref{fig:SED}.

In order to explain the observed dramatic photometric variability, 
we consider changes in the outburst properties of V346\,Nor,
or, alternatively, in the line-of-sight extinction.
We find that variable extinction alone is not sufficient to explain the 
observed spectral slope, as the change in colour between the
  high and low-accretion phase is much too shallow to be consistent 
  with dust extinction.  Between 1997/98 and 2010 the $K$-band magnitude
  dropped by $\Delta K=5.3$ with remarkably blue colours of $\Delta(J-K)=2.1$ and $\Delta(R-K)=2.6$.
  Dust extinction would result in much redder colours,
  e.g.\ $\Delta(J-K)=7.8$ and $\Delta(R-K)=36.6$ for the small 
  ISM-like dust population (0.1\,$\mu$m Silicate grains)
  that was detected around V346\,Nor \citep{sch05}.
  We also tested scenarios with a large dust grain-population (e.g.\ $10\mu$m grains: $\Delta(J-K)=4.6$, $\Delta(R-K)=16.6$),
  with a power-law grain size distribution, and dust grain mixtures,
  but were not able to reproduce the observed very blue colours with extinction alone.
  This finding is also consistent with the fact that the Silicate absorption feature 
  does not exhibit any significant changes in our 2005+2006+2007+2015+2016 mid-infrared spectra
  compared to the outburst phase \citep{sch05,qua07}, indicating that the 
  line-of-sight extinction remained unchanged.

We propose that the observed fading event was caused by a
dramatic drop in accretion luminosity.
We estimate the corresponding change in accretion rate by 
integrating the short-wavelength SED covered by our observations ($\lambda \lesssim 25\,\mu$m)
during the outburst phase (1990s) and near the photometric minimum (2010).
A major uncertainty concerns the line-of-sight extinction, where \citet{aud14}
proposed $A_V>12$\,mag. 
This value is higher than some earlier estimates \citep[e.g.\ $5.6^{+4.1}_{-3.7}$\,mag,][]{con10}, 
but supported by our SINFONI data, where H$_2$ 1-0 line transitions detected in the
vicinity of V346\,Nor favour very high extinction values (details will be outlined in Kraus et al., in prep.).
Also, the J/H/K-band colours measured in the low-accretion phase are consistent 
with  those of a classical T\,Tauri star for extinction values of $A_{V}=20\pm5$\,mag.
Adopting the lower-limit value of $A_V=12$\,mag implies that the accretion luminosity and
mass accretion rate changed by two orders-of-magnitudes,
while $A_V=15$\,mag implies a change by $\sim 10^3$.

These estimates suggest that V346\,Nor switched from an FUor outburst 
(with $\dot{M} \sim 10^{-4}$\,M$_{\sun}$\,yr$^{-1}$) to a rather passive phase with 
accretion rates similar to those observed towards T\,Tauri stars 
($\sim 10^{-7}$\,M$_{\sun}$\,yr$^{-1}$).
Remarkably, in the last few years (2011-2015), the object has brightened again
at all near-infrared wavelengths, indicating that
the object has entered a new outburst event. 
This is the first time that an FUor has been observed to switched between 
active-passive-active phases, although further monitoring will be needed to determine 
whether the new eruption will reach again the extreme accretion state of
the 1980s/1990s.

The near-infrared variability is also accompanied by changes in the mid-infrared flux,
where we observe an anti-correlated variability.
For instance, during the near-infrared fainting event between 2004 and 2007
the 8-13\,$\mu$m mid-infrared continuum flux increased by $\sim50$\% between 2004 and 2007.
During the new near-infrared brightening event in 2015/2016 at near-wavelength wavelengths,
the mid-infrared flux decreased again by $\sim70$\%.
This anti-correlated behaviour might provide important clues about how the 
intermediate disc regions responsed to the dramatic change in accretion luminosity.

\section{Discussion and Conclusions}
\label{sec:conclusions}

Our study shows that V346\,Nor has 
switched from an extreme FUor-type outburst into a much less active phase,
consistent with a change in the accretion rate 
by 2-3 orders of magnitude.
With ${\Delta}R=10.9$\,mag and ${\Delta}K=5.8$\,mag,
the observed dimming event far exceeded the variability 
that has been observed on other FUors before 
\citep[e.g.\ ${\Delta}R\approx3$\,mag, V899\,Mon;][]{nin15}.

Therefore, V346\,Nor opens for the first time the opportunity to 
study an FUor system in post-outburst and to investigate how the 
disc adjusted to the sudden drop in accretion luminosity.
For instance, some valuable insights might arise from modelling
the observed anti-correlated variability at near-infrared/mid-infrared wavelengths.
Also, the short-wavelength flux (R-band) started to decay a few year
ahead of the near-infrared flux (e.g.\ J/H/K-band; Fig.~\ref{fig:photometry}, top),
which might indicate that the disc needed a few years to react
to the sudden drop in accretion luminosity.

With an outburst duration of 20...25\,years,
the V346\,Nor outburst was shorter than expected for FUors 
($10^{2}$...$10^{3}$\,years), but significantly longer 
than for EXor (up to a few years).
This suggests that eruptive stars might not be well-represented 
by the classical bimodal FUor/EXor classification scheme, 
but exhibit a more continuous range of properties.
Already a few other objects with ``intermediate'' outburst durations
of a few years have been identified \citep{aud14,con16}
and started to blur the conventional distinctions between the FUor/EXor class.
However, V346\,Nor represents the most extreme case,
in particular as the object entered already into a new outburst,
less than a decade after the decay of its FUor-type eruption.
This recurrence cycle is orders-of-magnitudes shorter than
the typical time span of $\sim$5000...50,000\, years that has been
proposed for classical, isolated FUor outbursts \citep{sch13}, 
but might be consistent with the {\it clustered luminosity bursts}
that have been predicted by recent disc gravitational instability and fragmentation models
for the earliest stages of protostellar evolution \citep[e.g.][]{vor15}.

Besides its atypical lightcurve, V346\,Nor is also ambiguous in its spectral diagnostics,
as it features Br$\gamma$ and the CO 2.3\,$\mu$m bandheads in emission 
\citep{rei97a}, which is more commonly observed in EXors than FUors \citep[e.g.][]{lor06}.

With its unique characteristics concerning outburst duration,
repetition frequency, and spectroscopic diagnostics,
V346\,Nor provides important new insights on the relation between the
FUor and EXor phenomenon and might help to identify their triggering mechanism(s).

\section*{Acknowledgements}

We acknowledge support from an ERC Starting grant (Grant Agreement No.\ 639889),
STFC Rutherford fellowship and grant (ST/J004030/1, ST/K003445/1), a NASA Sagan fellowship,
Science Foundation Ireland grant (13/ERC/I2907), and
NSF grant (AST\,1311698).

%%%%%%%%%%%%%%%%%%%%%%%%%%%%%%%%%%%%%%%%%%%%%%%%%%



\bibliographystyle{mnras}

\vspace{-2mm}

%%%%%%%%%%%%%%%%%%%%%%%%%%%%%%%%%%%%%%%%%%%%%%%%%%

%%%%%%%%%%%%%%%%% APPENDICES %%%%%%%%%%%%%%%%%%%%%

%\appendix



%%%%%%%%%%%%%%%%%%%%%%%%%%%%%%%%%%%%%%%%%%%%%%%%%%

% Don't change these lines
\bsp	% typesetting comment
\label{lastpage}
\end{document}